\documentclass{article}

\usepackage{subcaption}
\usepackage[pdftex]{graphicx}
\usepackage{amsmath}
\usepackage{amssymb}
\usepackage{algorithm2e}

\begin{document}

\title{A circular parameterization\\for multi-sided patches}

\author{P\'eter Salvi\\{\small Budapest University of Technology and Economics}}

\date{2022}

\maketitle

\begin{abstract}
Most genuine multi-sided surface representations depend on a 2D
domain that enables a mapping between local parameters and global
coordinates. The shape of this domain ranges from regular polygons to
curved configurations, but the simple circular domain---to the best
of our knowledge---has not been investigated yet.

Here we fill this gap, and introduce a parameter mapping ideal for use
with periodic boundaries. It is based on circular arcs and satisfies
constraints often needed in actual surface formulations. The proposed
method is demonstrated through a corner-based variant of Generalized
B\'ezier patches.
\end{abstract}

\section{Introduction}

The natural design of many models, from simple household objects to
car bodies, often requires the use of multi-sided (non-quadrilateral)
surface patches. This is generally solved either by creating a larger
four-sided surface and cutting off the irrelevant parts (\emph{trimming}),
or by \emph{splitting} the multi-sided region into smaller, four-sided
subpatches.

Both of these approaches have their drawbacks; the alternative is to
use genuine multi-sided patches. These non-standard surfaces can be
defined by smoothly interpolating boundary constraints
(\emph{transfinite interpolation}), or by a network of control points.

Either way, a domain is needed. Although in some
cases~\cite{Sabin:1983,Zheng:1997} the domain does not take an explicit
geometric form, it is usually a 2D shape. In the beginning it was
assumed to be a regular $n$-sided polygon~\cite{Charrot:1984}, but
later this gave place to arbitrary convex polygons~\cite{Varady:2011},
concave polygons~\cite{Salvi:2018}, and even general curved
domains~\cite{Varady:2020}.

The rationale behind this progression is that a domain similar to the
3D configuration of the boundary curves reduces distortion. It is
especially important to mimic the angle at the vertices where boundary
curves meet. When this is equal to $\pi$, i.e., when the boundaries
have parallel tangents, most existing methods cannot be used (but
cf.~splits~\cite{Varady:2020} and periodic ribbons~\cite{Vaitkus:2021}).

In this paper we investigate a neglected option that seems tailored to these cases,
the \emph{circular} domain, and define a parameter mapping suitable
for use in many multi-sided patch formulations.

\section{Preliminaries}

Parameterizations of 2D domains come in many varieties; in the context
of multi-sided surfacing, these are used to map quadrilateral interpolants
onto the domain. Depending on the patch representation, there may be
several constraints on the parameters, as will be explained below.

The parameterization we are going to discuss in this paper is a
distance or \emph{height} parameter, usually denoted by $h$. It
measures the distance of an interior point from a base side (see
Figure~\ref{fig:conventional}), and needs to satisfy the following
properties:
\begin{figure}
  \centering
  \includegraphics[width = .35\columnwidth]{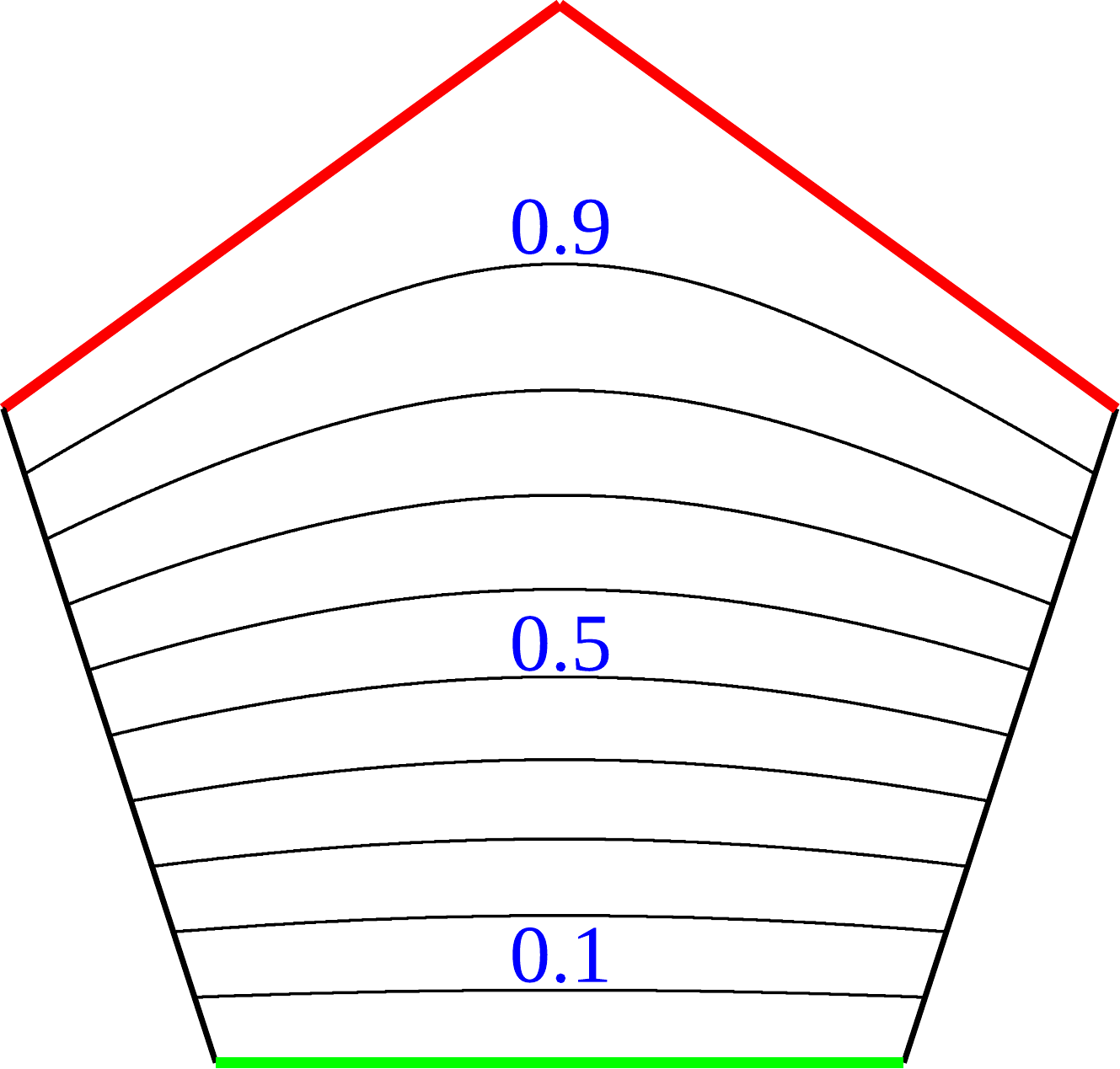}
  \caption{Constant parameter lines of a typical height parameter over
    a 5-sided polygonal domain. The base side is shown in green; distant sides in red.}
  \label{fig:conventional}
\end{figure}

\begin{enumerate}
\item $h=0$ on the base side.
\item $h$ is continuous and varies monotonically.
\item $h$ changes uniformly from $0$ to $1$ on the sides adjacent to the base side.
\end{enumerate}

\noindent Such a mapping can be created for all possible base sides, resulting
in $n$ mappings, denoted by $h_i$ ($i=1\dots n$). Depending on the
application, there are often other constraints:

\begin{enumerate}
  \setcounter{enumi}{3}
\item $h\leq1$ everywhere inside the domain.
\item $h=1$ on all distant sides, i.e., on all sides not equal or
  adjacent to the base side. [\emph{full mapping}]
\item $h_{i-1}'=-h_{i+1}'$ on the $i$th side. [\emph{constrained mapping}]
\end{enumerate}

\noindent It is easy to see that (5) subsumes (4).

There have been parameterizations over other domains satisfying some of the
above constraints;
Salvi et al.~\cite{Salvi:2017:WAIT}, in particular, showed how a
mapping can be constructed that has all of these properties: there the
parameterization itself is a (singular, 1D) multi-sided
patch based on the work of Kat\=o~\cite{Kato:1991}.

Here we propose a height parameter that also satisfies all the
constraints mentioned above, but over a circular domain.

\section{Circular parameterization}

A circular domain is, well, a circle---for ease of computation, let it
be the unit circle around the origin. We allocate equal arcs to each side. Without loss
of generality, let us assume that the base side is the
$[-\pi/n,\pi/n]$ arc.

\begin{figure}[!hb]
  \centering
  \includegraphics[width = .55\columnwidth]{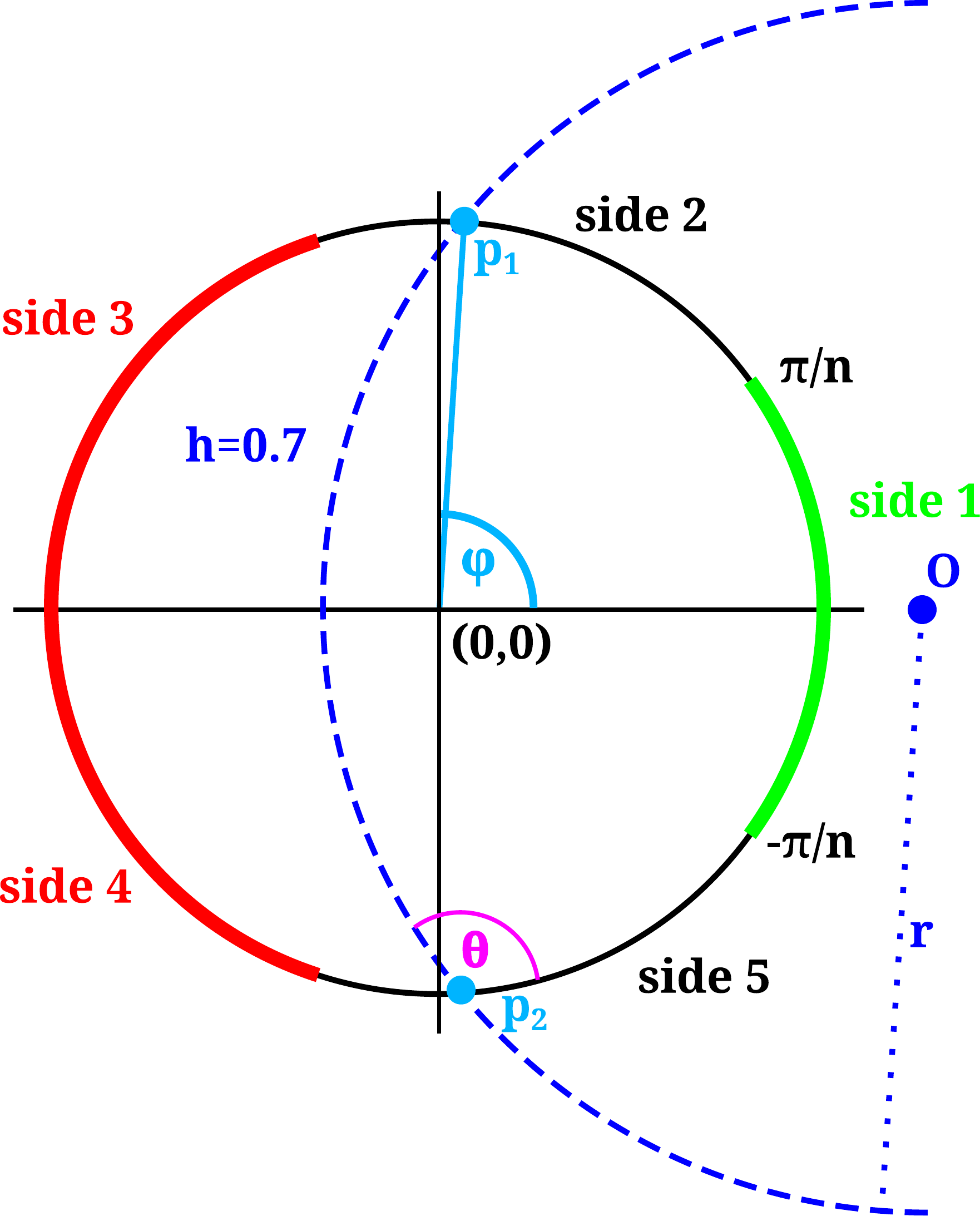}
  \caption{Notations in the canonical position, shown on a 5-sided configuration.
  Side 1 (green) is the base side, while sides 3 and 4 are the distant sides.}
  \label{fig:explanation}
\end{figure}

First we tackle the inverse problem, i.e., defining the constant
parameter line for a given $h$ value. We will use circular arcs for
that purpose, with endpoints $\mathbf{p}_1$ and $\mathbf{p}_2$ uniformly ranging over
the adjacent sides (property 3):
\begin{equation}
  \mathbf{p}_1 = (\cos\varphi,\sin\varphi),\quad
  \mathbf{p}_2 = (\cos\varphi,-\sin\varphi)
\end{equation}
where $\varphi=(2h+1)\pi/n$, see Figure~\ref{fig:explanation}.

\newpage
The idea is that we determine the circular arc by its angle to the
domain circle ($\theta$), which we require to change from $0$ to $\pi$
as $h$ progresses from $0$ to $1$, i.e., $\theta=h\pi$. This means that the arc will be
part of the domain circle when $h=0$ or $h=1$. A constant parameter
line defined like this obviously satisfies all properties,
including~(5) and~(6).

Straightforward algebra leads to the equations for the center point and
radius of the constant parameter line:
\begin{equation}
  \label{eq:Or}
  \mathbf{O} = \left(\frac{\sin\theta}{\sin\psi},0\right),\quad
  r = \left|\frac{\sin\varphi}{\sin\psi}\right|,
\end{equation}
where $\psi = \theta-\varphi = h\pi-\varphi$.
(For details, see Appendix~\ref{app:proof}.)

\subsection{Mapping algorithm}

Mapping a $\mathbf{p}=(u,v)$ point in the circle to a $h$ parameter unfortunately
leads to an equation of degree $2n$. Instead, we propose a procedural
algorithm using bisection. It uses the fact that points with the
coordinate $\hat{u}=\cos(\pi/(n-2))$ are on a \emph{straight} constant
parameter line (a circle with infinite radius), associated with
$\hat{h}=1/(n-2)$.

\begin{algorithm}
  \SetKwFunction{bisection}{bisection}
  \uIf{$u>\hat{u}$}{\Return \bisection{$\Delta$,\,$0$,\,$\hat{h}-\epsilon$}}
  \uIf{$u<\hat{u}$}{\Return \bisection{$\Delta$,\,$\hat{h}+\epsilon$,\,$1$}}
  \Return $\hat{h}$
\end{algorithm}

\noindent
Here {\ttfamily bisection} searches for a value between its second and third
arguments where the deviation vanishes. The deviation is given as the
function $\Delta(h)$ shown below:
\begin{equation}
  \Delta(h)=\|\mathbf{p}-\mathbf{O}(h)\|-r(h).
\end{equation}
This is a fast and robust algorithm, which is very easy to implement;
see Appendix~\ref{app:height}.

\subsection{Examples}

\begin{figure}

  \begin{subfigure}{.4\columnwidth}
    \includegraphics[width = \textwidth]{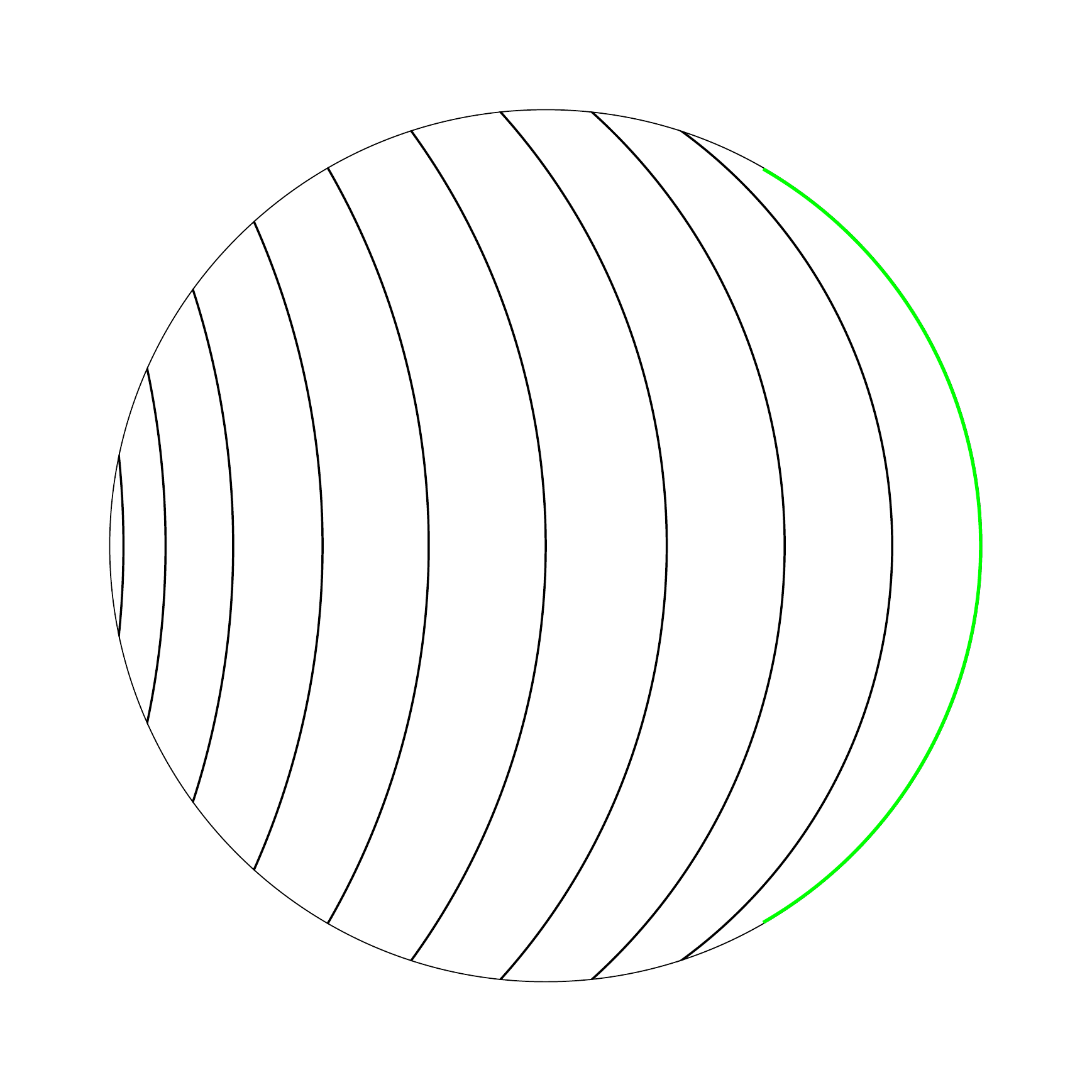}
    \caption{$n=3$}
    \label{fig:ex3}
  \end{subfigure}
  \hfill
  \begin{subfigure}{.4\columnwidth}
    \includegraphics[width = \textwidth]{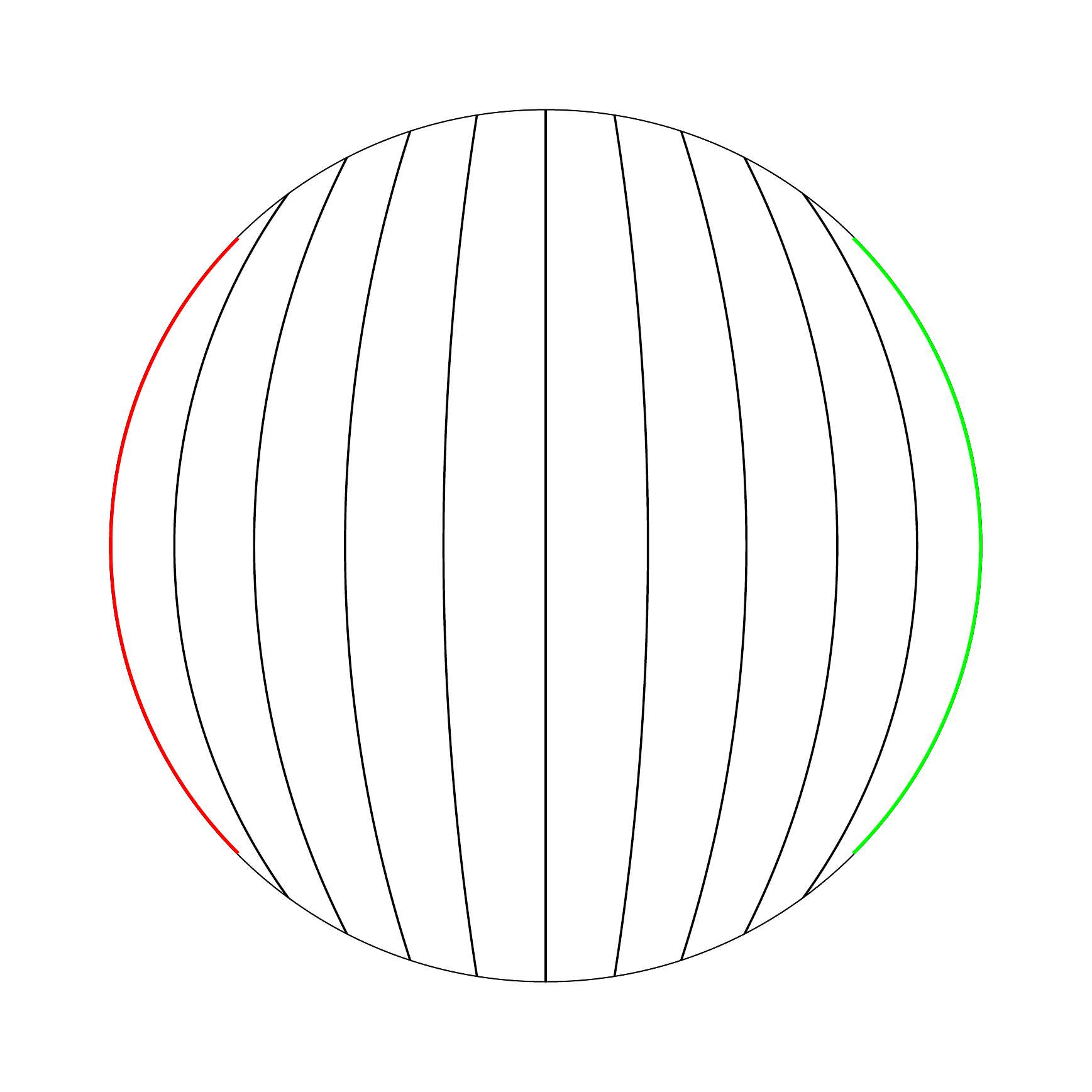}
    \caption{$n=4$}
    \label{fig:ex4}
  \end{subfigure}

  \begin{subfigure}{.4\columnwidth}
    \includegraphics[width = \textwidth]{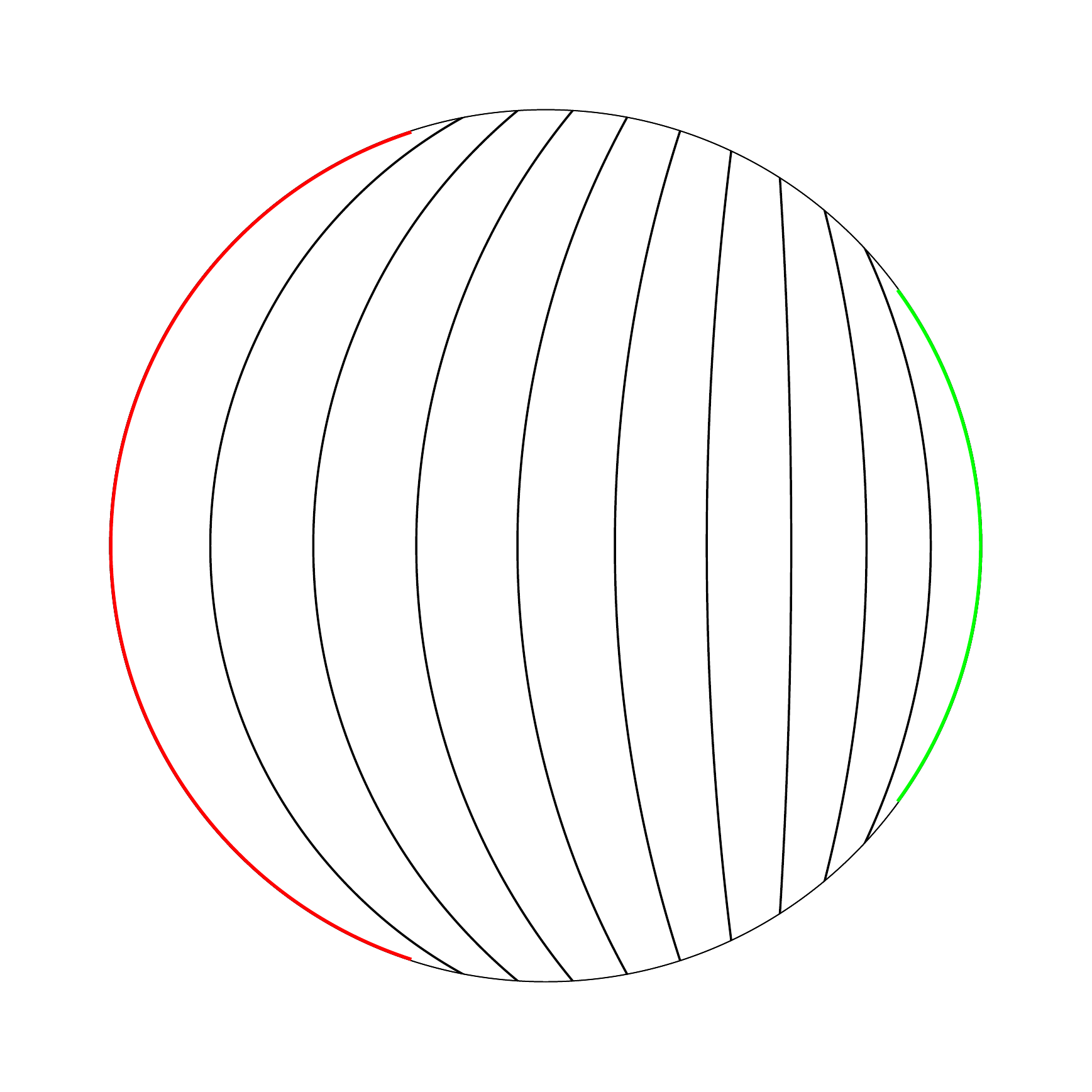}
    \caption{$n=5$}
    \label{fig:ex5}
  \end{subfigure}
  \begin{subfigure}{.4\columnwidth}
    \includegraphics[width = \textwidth]{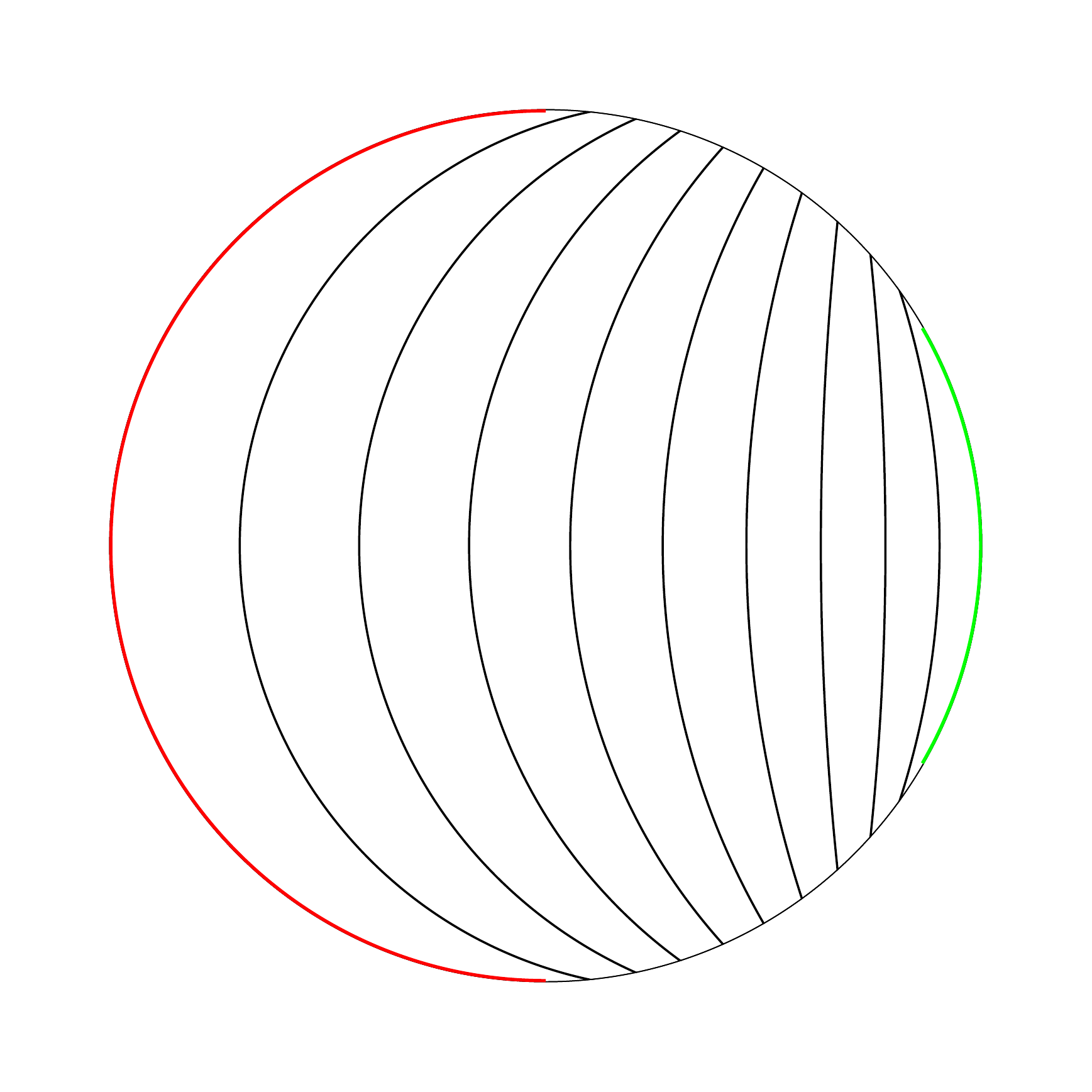}
    \caption{$n=6$}
    \label{fig:ex6}
  \end{subfigure}
  \begin{subfigure}{.4\columnwidth}
    \includegraphics[width = \textwidth]{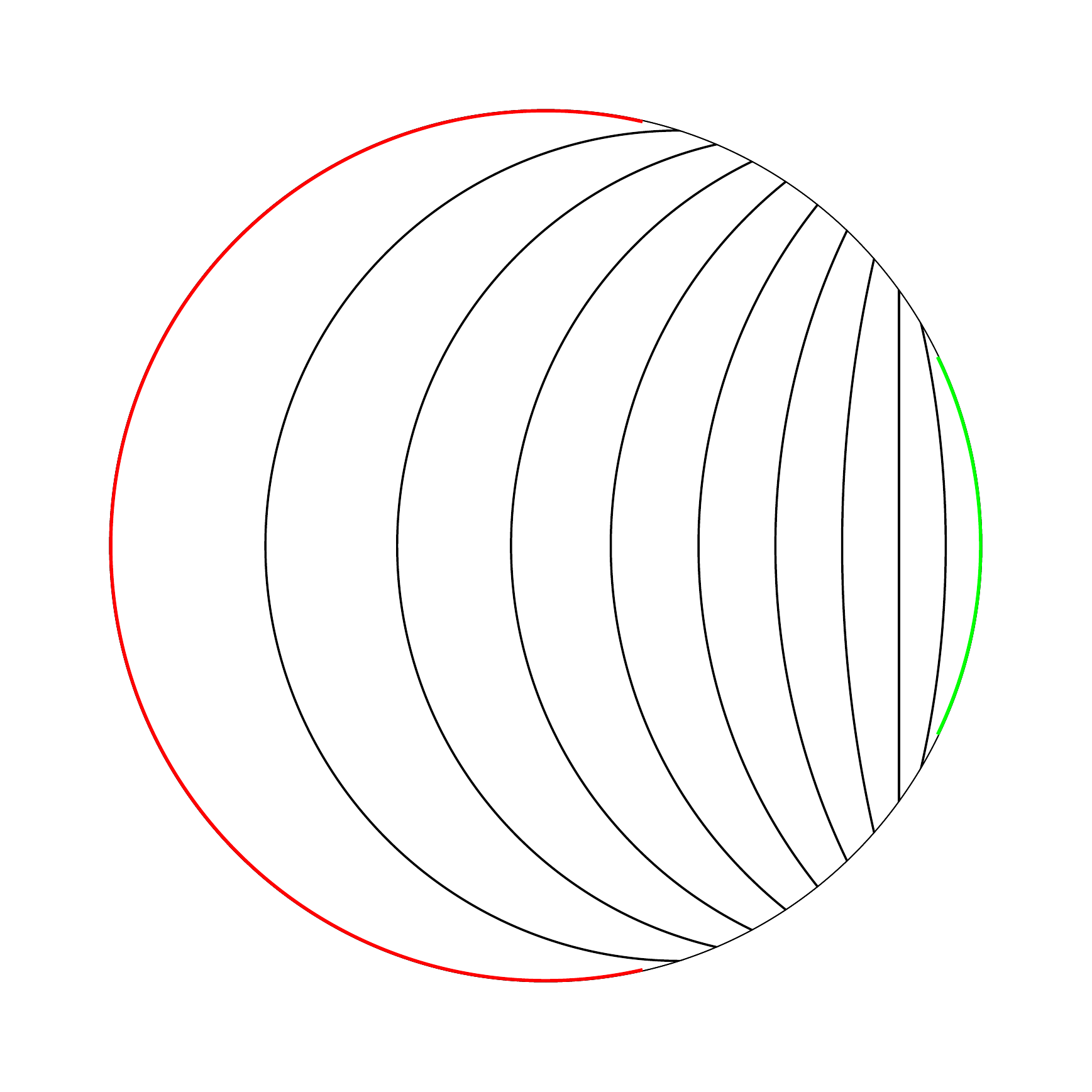}
    \caption{$n=7$}
    \label{fig:ex7}
  \end{subfigure}
  \hfill
  \begin{subfigure}{.4\columnwidth}
    \includegraphics[width = \textwidth]{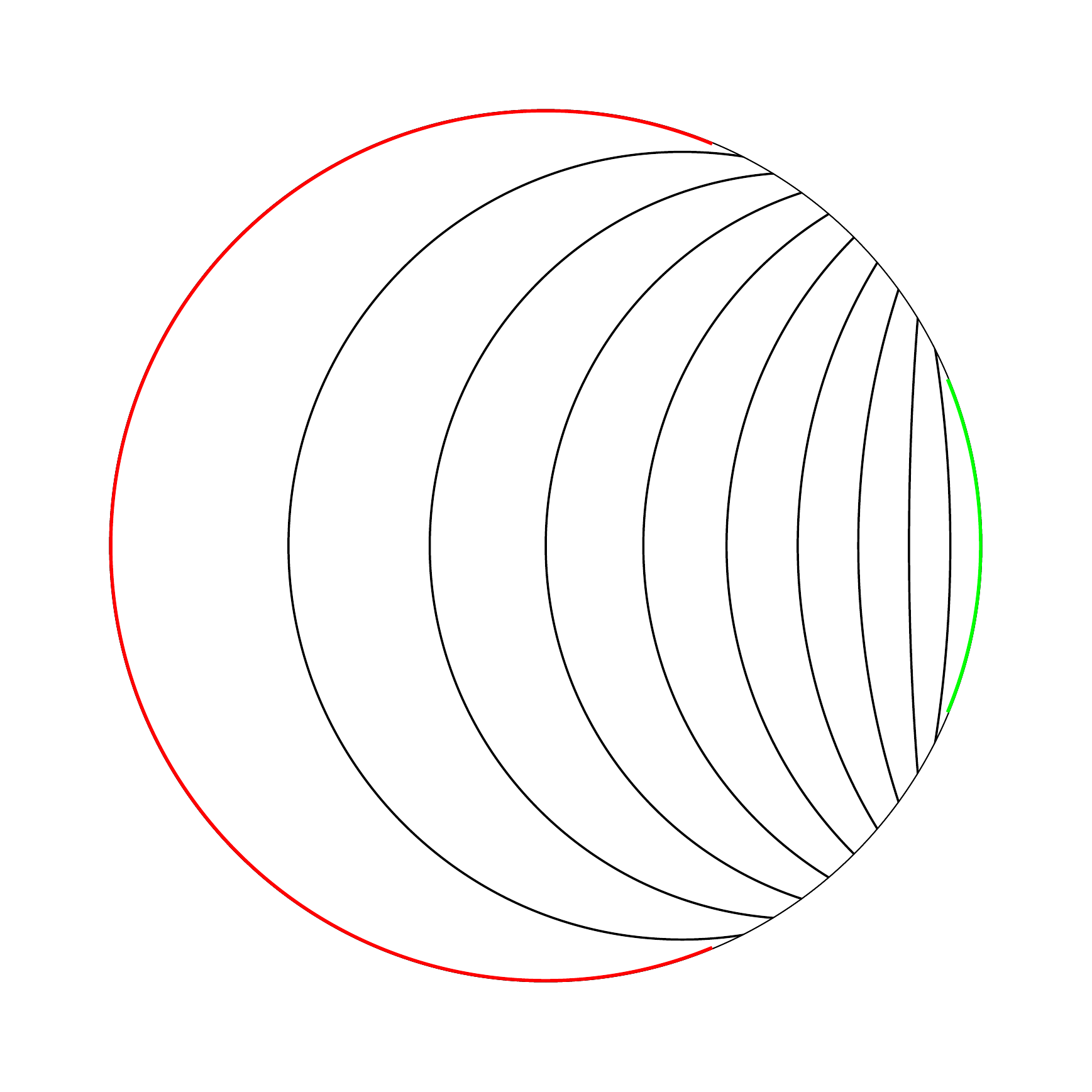}
    \caption{$n=8$}
    \label{fig:ex8}
  \end{subfigure}

  \caption{Constant parameter lines on domains with 3 to 8
    ``sides''. The base side is shown in green; distant sides in red.}
  \label{fig:examples}
\end{figure}

\begin{figure}
  \centering
  \includegraphics[width = .35\columnwidth]{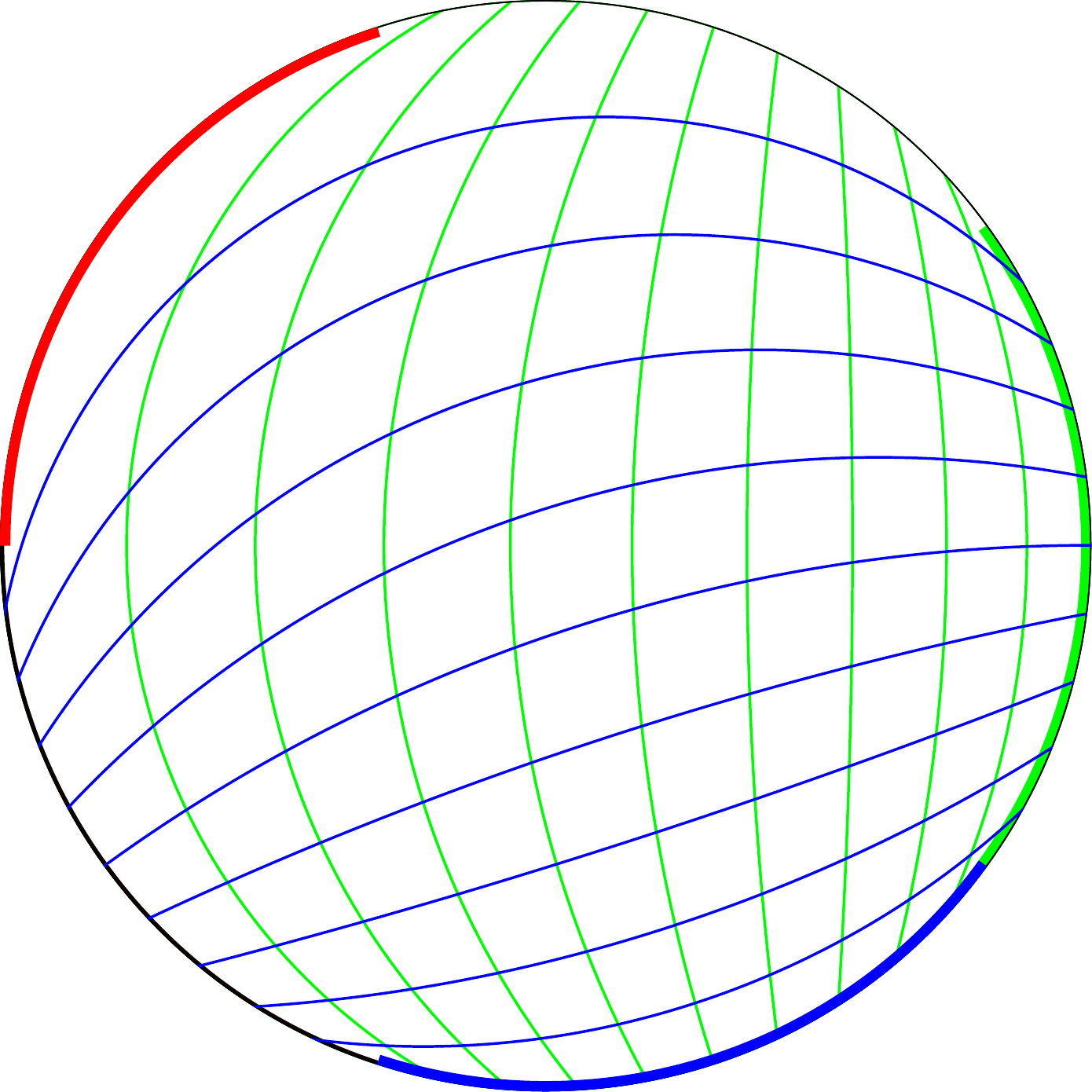}
  \caption{Corner parameterization for a 5-sided patch. The base sides and their respective
    constant parameter lines are shown in green and blue; the distant side is shown in red.}
  \label{fig:corner}
\end{figure}

Parameterizations of 3- to 8-sided domains are shown in
Figure~\ref{fig:examples}. Taking parameterizations associated with
adjacent base sides (i.e., $h_{i-1}$ and $h_i$), we get a bivariate
mapping (Figure~\ref{fig:corner}), suitable for the parameterization
of corner interpolants, as will be demonstrated in the next section.

\begin{figure}
  \centering
  \includegraphics[width = .35\columnwidth]{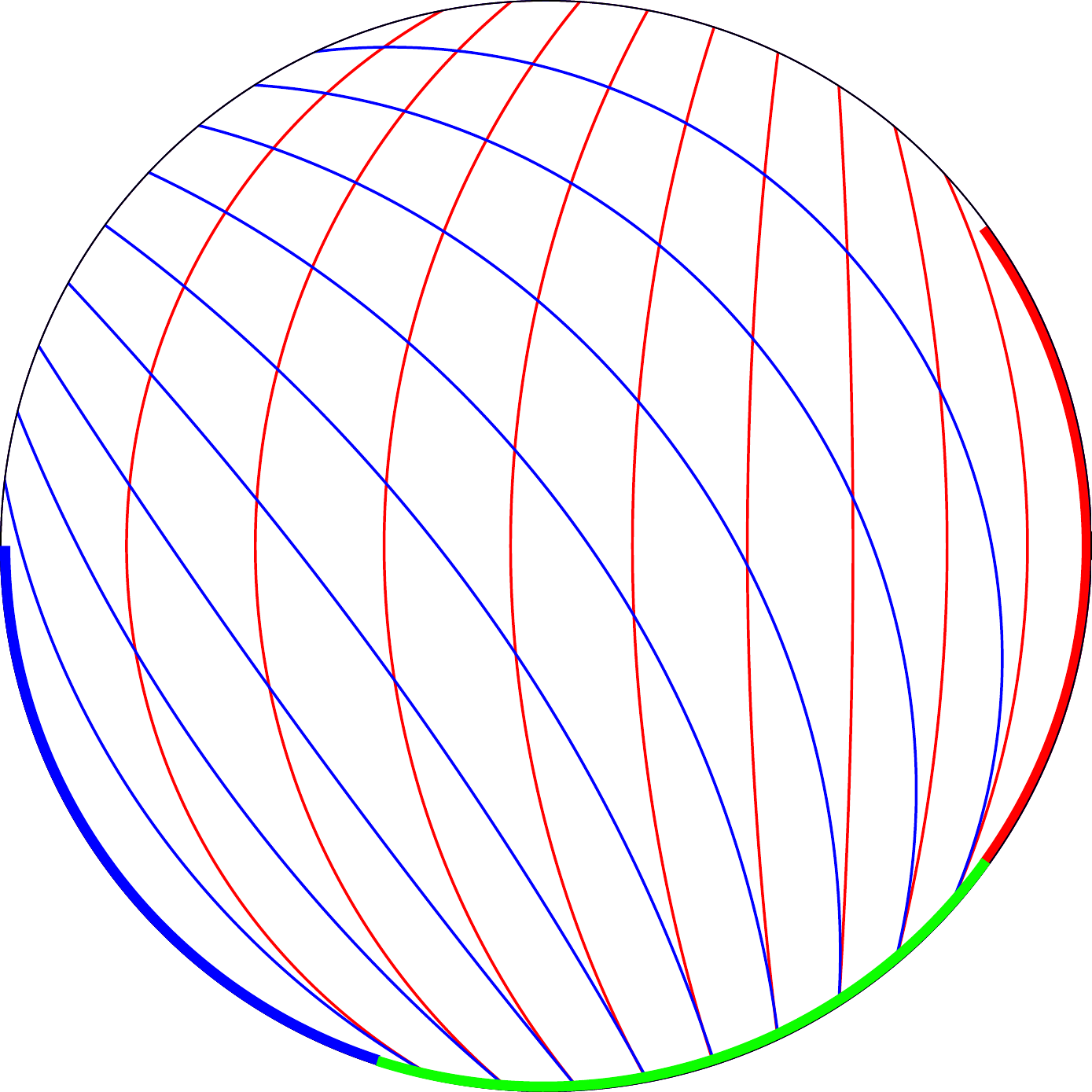}
  \caption{Demonstration of the derivative constraint on a 5-sided patch.
    The base sides and their respective constant parameter lines are shown in red and blue;
    the lines start in the same direction near the green arc.}
  \label{fig:constraint}
\end{figure}

The derivative constraint is exemplified in
Figure~\ref{fig:constraint}, where the constant parameter lines of
$h_{i-1}$ and $h_{i+1}$ have the same tangent at the $i$th side.

\section{Application to Overlap-GB (OGB) patches}

The Overlap patch~\cite{Varady:1991} is a corner-based formulation,
where the surface is defined as the sum of corner interpolants. Each
interpolant is given by its corner vertex, two derivatives and a twist
vector. The parameterization used in the original paper was a
$G^1$-continuous composite function, but it can be replaced by our
circular mapping.

Generalizing the Overlap patch to arbitrary degrees leads to a
formulation very similar to the Generalized B\'ezier (GB)
patch~\cite{Varady:2016:EG}, but without the rational weight functions.
Since patch formulations are not the topic of this paper, we only show
the required equations here:
\begin{equation}
  S=\sum_{i=1}^n\sum_{j=0}^{\lfloor d/2\rfloor}\sum_{k=0}^{\lfloor d/2\rfloor}
  \mathbf{P}_{ijk}B^d_j(h_{i+1})B^d_k(h_i)+\mathbf{P}_0B_0,
\end{equation}
where $\mathbf{P}_{ijk}$ are points constituting a control network, $B^d_\ell(h)$
are degree-$d$ Bernstein polynomials, $\mathbf{P}_0$ is a central control
point, and $B_0$ is the weight deficiency:
\begin{equation}
  B_0 = 1 - \sum_{i=1}^n\sum_{j=0}^{\lfloor d/2\rfloor}\sum_{k=0}^{\lfloor d/2\rfloor}
  B^d_j(h_{i+1})B^d_k(h_i).
\end{equation}
Here the degree $d$ is assumed to be odd; it is easy to modify the
above equations to allow even values, as well.

\begin{figure*}[!ht]
  \centering
  \includegraphics[width = .43\textwidth]{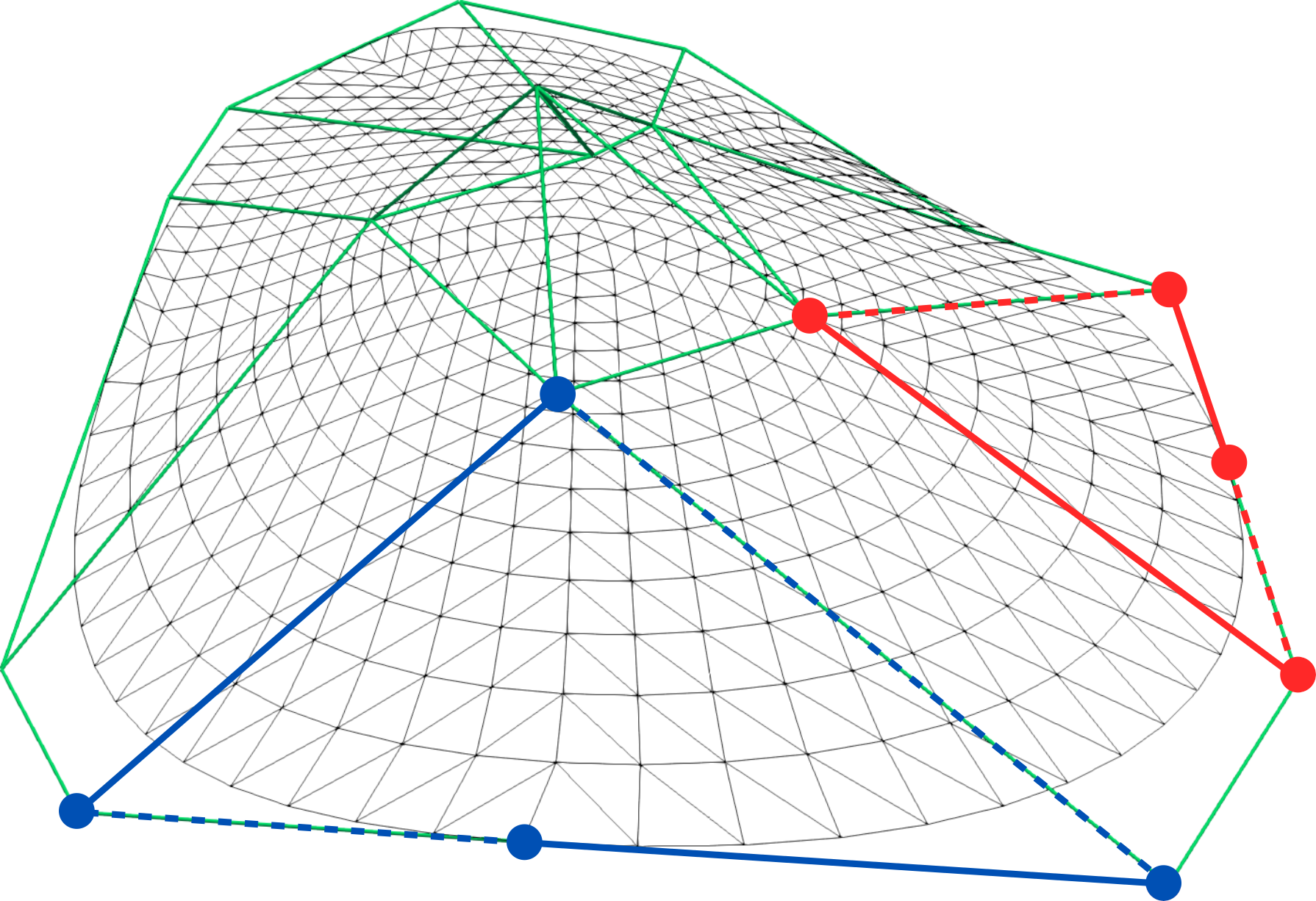}\hfill
  \includegraphics[width = .43\textwidth]{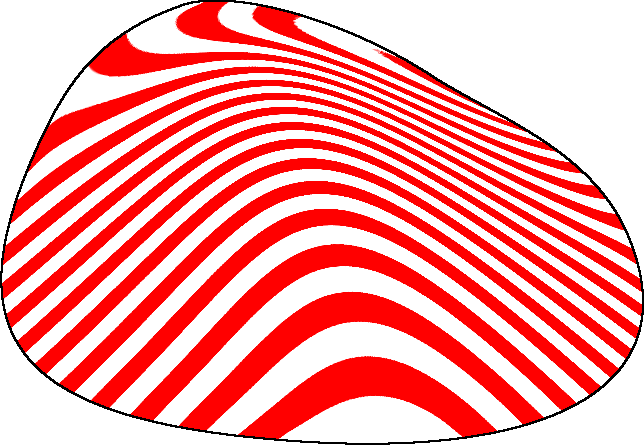}
  \caption{A 5-sided cubic OGB patch (\emph{left}: control network highlighting two corner
    interpolants; \emph{right}: isophote lines).}
  \label{fig:ogb}
\end{figure*}

Note that interpolation of the boundary constraints requires a full,
constrained $h$-mapping. Figure~\ref{fig:ogb} shows a 5-sided cubic patch
based on the above equations, using circular parameterization. From
the tessellation it can be seen that no distortion has been introduced,
and the isophote lines also flow naturally.

\section*{Conclusion}

We have investigated the simple circle as a candidate for a
multi-sided domain, along with a distance parameter mapping satisfying
all properties commonly needed for use in $n$-sided patches.

The proposed parameterization is perfect for handling periodic
boundaries, and while it cannot be evaluated explicitly, it is both
conceptually and algorithmically simpler than previous
solutions~\cite{Salvi:2017:WAIT,Vaitkus:2021}.

Future work includes the investigation of derivative behavior in the
corners.

\section*{Acknowledgements}

The author would like to thank Tam\'as V\'arady and M\'arton Vaitkus for
the many fruitful discussions on this topic.

\bibliographystyle{plain}
\bibliography{circular.bib}

\pagebreak

\appendix

\section{Proof of Eq.~(\ref{eq:Or})}
\label{app:proof}

The tangent to the unit circle at $\mathbf{p}_2=(\cos\varphi,-\sin\varphi)$ is
\begin{equation}
  \mathbf{t}=(\sin\varphi,\cos\varphi),
\end{equation}
which we rotate by the angle $\theta$ to get the tangent to the
constant parameter arc:
\begin{align}
  \mathbf{d}&=
  (\sin\varphi,\cos\varphi)
  \begin{pmatrix}\phantom-\cos\theta&\sin\theta\\-\sin\theta&\cos\theta\end{pmatrix}\nonumber\\
  &=(\sin(\varphi-\theta),\cos(\varphi-\theta)).
\end{align}
This is perpendicular to the radius from $\mathbf{O}=(x,0)$, i.e.,
\begin{equation}
  \langle \mathbf{p}_2 - \mathbf{O}, \mathbf{d}\rangle = 0,
\end{equation}
which leads to
\begin{equation}
  (\cos\varphi-x)\sin(\varphi-\theta)=\sin\varphi\cos(\varphi-\theta).
\end{equation}
Using the fact that
\begin{equation}
  \sin\varphi\cos(\varphi-\theta)-\cos\varphi\sin(\varphi-\theta)=\sin\theta,
\end{equation}
we arrive at
\begin{equation}
  x=-\frac{\sin\theta}{\sin(\varphi-\theta)}=\frac{\sin\theta}{\sin(\theta-\varphi)}
    =\frac{\sin\theta}{\sin\psi}.
\end{equation}
The radius can be computed by
\begin{align}
  r&=\|\mathbf{p}_1-\mathbf{O}\|=\left\|\left(\cos\varphi-\frac{\sin\theta}{\sin\psi},\sin\varphi\right)\right\|\nonumber\\
  &=\sqrt{\sin^2\varphi+\cos^2\varphi-2\cos\varphi\frac{\sin\theta}{\sin\psi}+\frac{\sin^2\theta}{\sin^2\psi}}\\
  &\text{(using $\theta=\varphi+\psi$)}\quad
  =\sqrt{\frac{\sin^2\varphi}{\sin^2\psi}}=\left|\frac{\sin\varphi}{\sin\psi}\right|.~\square\nonumber
\end{align}

\pagebreak

\begin{figure}[!ht]
  \centering
  \includegraphics[width = .4\columnwidth]{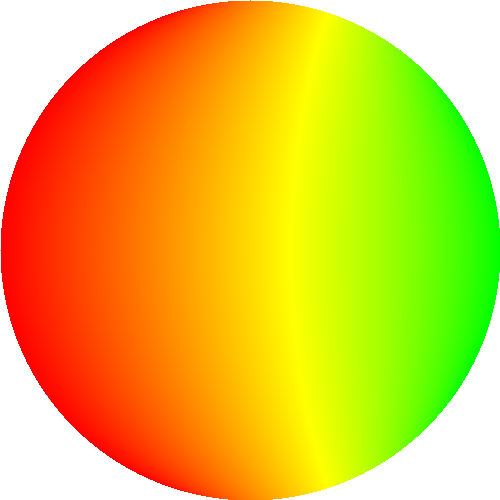}
  \caption{A bitmap showing $h$-values on a 5-sided domain
    (green: $h=0$, yellow: $h=0.5$, red: $h=1$).}
  \label{fig:bitmap}
\end{figure}

\section{Computation of the height map}
\label{app:height}

The following \textsc{Julia} program computes the $h$ mapping of a given point,
which is first rotated to the canonical position, i.e., where the base side
is the $[-\pi/n,\pi/n]$ arc.

\bigskip

\noindent
\includegraphics[width = .55\columnwidth]{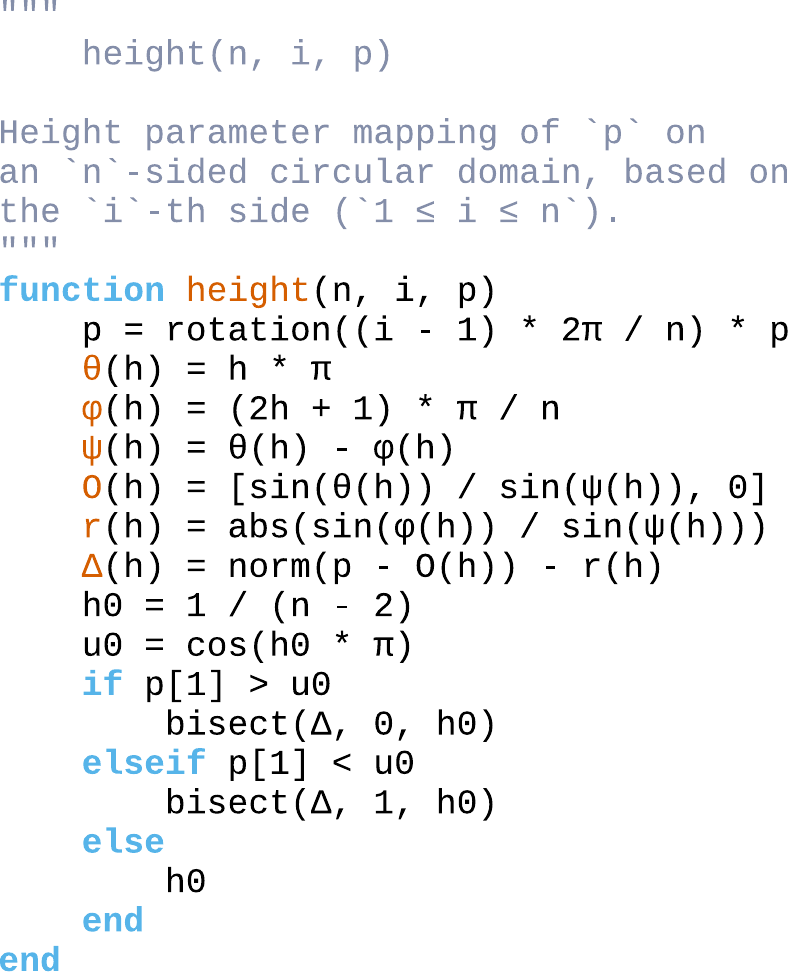}

\bigskip

\noindent
Note that the above code needs the \texttt{LinearAlgebra} module of
the standard library. The bisection algorithm is implemented as below.

\noindent
\includegraphics[width = .67\columnwidth]{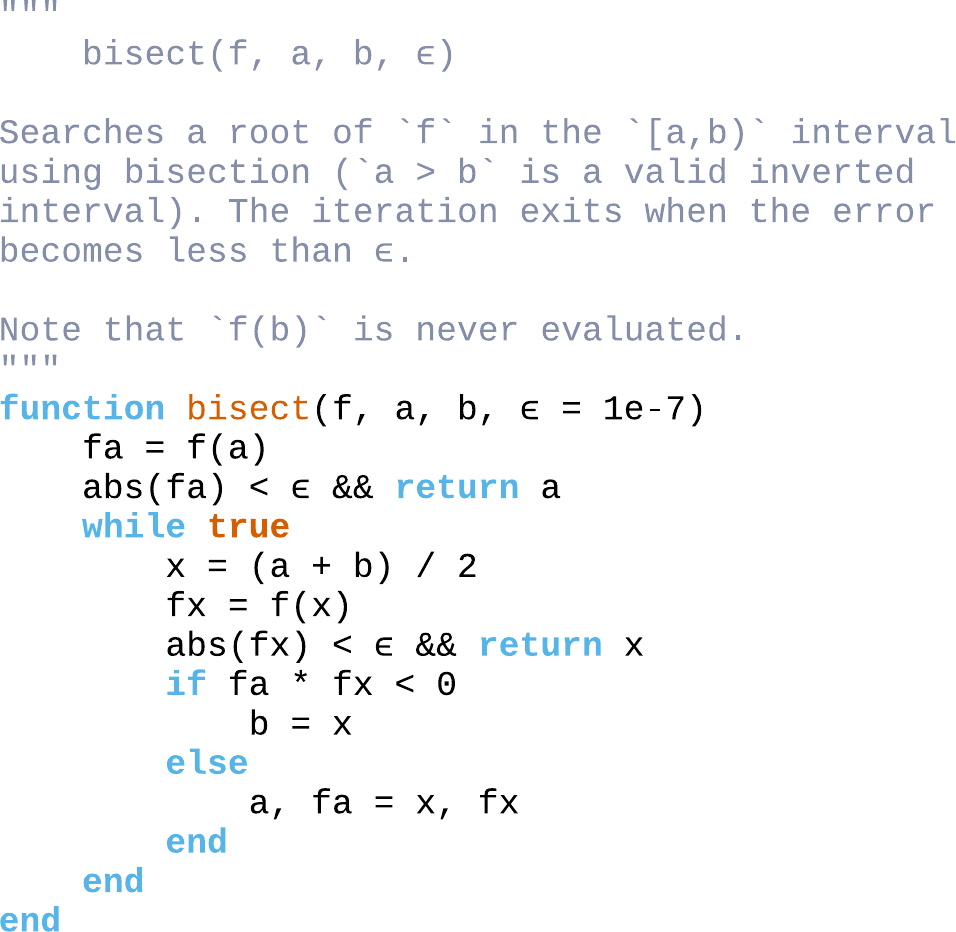}

\bigskip

\noindent
Finally, the rotation matrix is computed simply by:

\bigskip

\noindent
\includegraphics[width = .67\columnwidth]{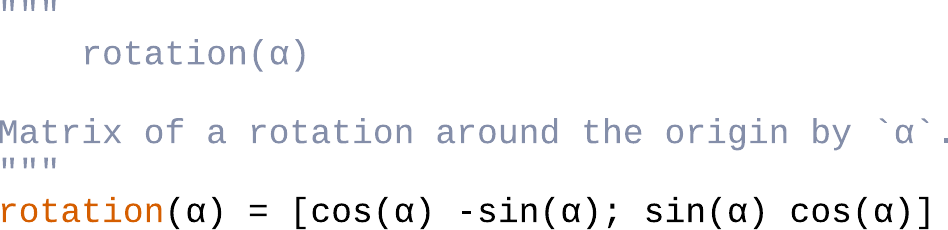}

\bigskip

\noindent
A bitmap generated by this code is shown in Figure~\ref{fig:bitmap}.

\end{document}